\documentclass{article}
\usepackage{amssymb}
\usepackage{amsmath}

\newcommand{\pa}{\partial}

\newcommand{\myref}[1]{(\ref{#1})}

\newcommand{\om}{\omega} 

\newcommand{\de}{\delta}

\newcommand{\al}{\alpha}
\newcommand{\eps}{\varepsilon}
\newcommand{\la}{\lambda}

\newcommand{\be}{\beta}

\renewcommand{\leq}{\leqslant}

\newcommand{\lan}{\langle}
\newcommand{\ran}{\rangle}

\newcommand{\demi}{\frac{1}{2}}

\newcommand{\mcal}[1]{\mathcal{#1}}

\newlength{\somme}
\newlength{\sommep}
\newcommand{\intvide}{\rule[-\sommep]{0cm}{\somme}}

\newlength{\sommebis}
\newlength{\sommepbis}
\usepackage{amsmath}
\usepackage{amssymb}
\usepackage{graphics}
\usepackage{anysize}
\marginsize{2cm}{2cm}{2cm}{2cm}

\begin{document}
\title{The notion of persistence applied to breathers in thermal equilibrium}
\author{Jean Farago}

\date{\today}

\maketitle
\begin{abstract}
We study the thermal equilibrium of nonlinear Klein-Gordon chains at
the limit of small coupling (anticontinuum limit). We show that the
persistence distribution associated to the local energy density is a
useful tool to study the statistical distribution of so-called thermal
breathers, mainly when the equilibrium is characterized by long-lived
static excitations; in that case, the distribution of persistence
intervals turns out to be a powerlaw. We demonstrate also that this generic
behaviour has a counterpart in the power spectra, where the
high frequencies domains nicely collapse if properly rescaled. These
results are also compared to non linear Klein-Gordon chains with a
soft nonlinearity, for which the thermal breathers are rather mobile
entities. Finally, we discuss the possibility of a breather-induced
anomalous diffusion law, and show that despite a strong slowing-down
of the energy diffusion, there are numerical evidences for a normal
asymptotic diffusion mechanism, but with exceptionnally small diffusion coefficients.
\end{abstract}

\section{Introduction}
For a decade or so, a growing interest has been manifested for the
so-called breather modes \cite{flachwillis,aubry60}. Two factors have contributed mainly to this
popularity. First, these localized excitations are generic in
arrays of oscillators, provided that both nonlinearity and a certain amount of discreteness
are features of the system \cite{mackayaubry}; as such, they are a priori ubiquitous in
nature and have thus a certain universality. But it was shown also
that systems where breather modes can be found could display a strong
slowing down of the energy diffusion, leading apparently to anomalous
subdiffusive behaviours reminiscent to glassy states \cite{tsironisaubry}; it was
recognized that the slowing down is intimately related to the
existence of so-called pinned breathers \cite{lindenberg,eleftheriou,ivanchenko}. The difficulty however
has remained up to now to bind these two features beyond qualitative
arguments, due to the fact that, in contrast with topologic solitons for instance,
the breather modes do not possess a true identity as dynamical
elementary excitations, what makes their very existence 
problematic in realistic situations where systems are at or close to
equilibrium. In these cases, a ``breather'' stricto sensu does no
longer exist, even if clear dynamical features akin to them  are
undoubtedly observable. 

The aim of this paper is to show that the concept of
\textit{persistence distribution} can be a useful tool and give an
insight into the ``breathers'' distribution in a thermal
context. Several studies have been already devoted to that subject
\cite{lindenberg,eleftheriou,ivanchenko,eleftheriouflachtsironis}, but they all faced the
empirical nature of a breather at $k_BT\neq 0$, and were forced to
circumvent the problem by ad hoc recipes. The persistence
distribution allows to overcome this difficulty: instead of focusing
on the problematic recognition of breathers among a thermalized chain,
one integrates the time axis into the problem and study the trend a
site has to stay above the mean energy level for a time arbitrary
long, i.e. its ability to persist above (or below) its average. By
this way, we give up a real extraction of a breather distribution, but
catch really the main interesting property of these thermal ``breathers'', that is
their ability to make last abnormally an energy fluctuation.

For a stochastic
process $X(t)$, the persistence distribution $\mcal{P}_\pm(\tau)$ is defined as the
probability that the process stays above ($\mcal{P}_+$) or below
($\mcal{P}_-$)  its mean value during the \textit{whole} time interval
$[0,\tau]$. This relatively simple definition hides a
complicated observable, as $\mcal{P}_\pm(\tau)$ assumes the knowledge
of the whole dynamics from the time origin on\cite{majumdar}. We will
show that this concept is particularly adapted to systems where
thermal breathers are pinned, as they lead naturally to persistent
local energy density: actually, the asymptotic branch of the
persistence distribution can be taken as a definition of the thermal
breathers distribution (see below). Besides, for systems where mobile
thermal breathers are excited, the persistence distribution yields
naturally a measure of the lifetime of the static entity created by
the collisions of two breathers.

In this paper, we focus on discrete Klein-Gordon lattices described by
the Hamiltonian
\begin{align}\label{H}
  \mcal{H}&=\sum_i e_i\\
  e_i&=\demi\dot{x}_i^2+ V(x_i)+\frac{k}{2}x_i(2x_i -x_{i+1}-x_{i-1})\nonumber
\end{align}
The dynamics of the lattices is exclusively microcanonical, and the
temperature is extracted by the usual ensemble equivalence.
As we are interested in the role of thermal breathers, we study
systems which are close to the anticontinuum limit, that is
$k/\om_0^2\ll 1$, where $\om_0^2$ is defined through
$V(x)=C^t+\demi\om_0^2 (x-x_\text{min})^2+o(x-x_\text{min})^2$; in the
following we take $k/\om_0^2=0.1$.

As regards the on-site potentials $V(x)$, we have considered the hard
$\phi^4$ potential $V_1(x)=\demi x^2+\frac{1}{4}x^4$ (famous for its
pinned breathers), a soft potential
$V_2(x)=\frac{1}{3}\log\cosh(\sqrt{3}x)$ (with a Taylor expansion
$V_2(x)=\demi x^2-\frac{1}{4}x^4+o(x^4)$; this potential has typically
mobile thermal excitations) and the harmonic potential
$V_3(x)=\demi x^2$. Note that $\om_0^2=1$ in all these cases.

In a first part, we introduce the distribution of time intervals of
persistent energy excess (TIPEE) $P(t)$ (closely related to the persistence),
and discuss the numerical results for $P(t)$ for the three on-site
potentials. We then study the power spectrum associated to these
models and show some connections with the persistence
behaviours. Next, we discuss the possibility of an anomalous diffusion
for nonlinear Klein-Gordon with hard anharmonicity. Finally, we
introduce the phenomenological concept of breather lifetime, and show
that for static breathers of moderate-to-high energies, these
lifetimes are exponentially increasing with their energy.

\section{The TIPEE and persistence distributions}

The time interval of persistent energy excess $P(t)$ is defined for
positive and negative $t$. For positive $t$ (resp. negative), it is the distribution function
of the time intervals during which a given site has its energy $e_i$ greater
(resp. lower) than the mean local energy $\lan e_i\ran$. It is very
simple to measure, but actually this is a rather complicated
statistical concept, just like the persistence distributions
\cite{majumdar}. By the way, $P(t)$ is closely related to the
persistence distributions above and below
the mean energy by the relations (see Appendix)
$\mcal{P}_\pm(\tau)=\int_\tau^\infty dt
P(\pm t)(t-\tau)/\int_{0}^\infty dt\ tP(\pm t)$. In the litterature
of stochastic processes, the persistence distribution is  widely
used, but actually for our purpose, the TIPEE distribution $P(t)$ is
more appropriate : we'll focus on it henceforward.

To get $P(t)$ we must compulsorily perform numerical simulations : it
is an object so complicated from the statistical point of view, that
it is likely that no exact calculation would be amenable even for the
integrable harmonic chain $V(x)=V_3(x)$. 

\subsection{Harmonic chain $V(x)=V_3(x)$}

\begin{figure}[h]
\centerline{  \resizebox{14cm}{!}{\includegraphics{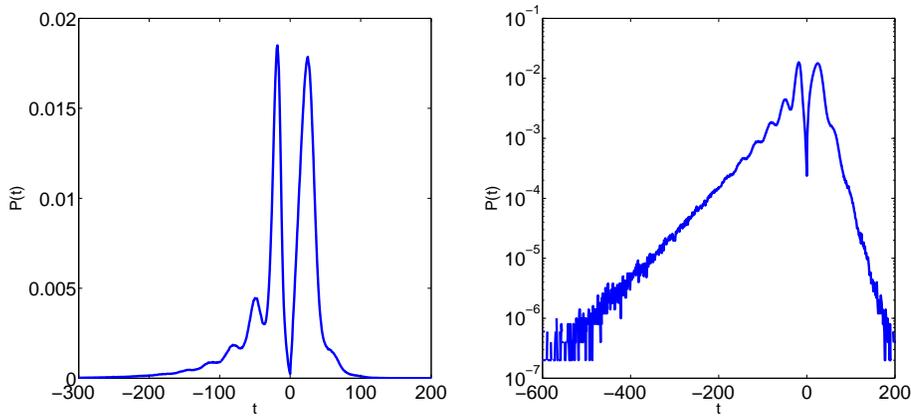}}}
\caption{Distribution $P(t)$ for the harmonic potential $V_3(x)=\demi
  x^2$. Left : linear-linear plot; right : semilog plot.}\label{TIPEE_V3}
\end{figure}

The numerical result for
$P(t)$, plotted in figure \ref{TIPEE_V3},
 is independent of the
temperature, due to the invariance of the dynamics under a dilatation
$(x,v)\rightarrow (\la x,\la v)$. The two main peaks that characterize the
distribution give the typical times of a fluctuation; one easily
verifies that they are of the same order of magnitude as $1/k$: a wave packet travels at $v_g$
throughout the system and is likely to create a positive fluctuation
at a given site for a time or order $1/v_g\sim 1/k$. 
As regards the tails of the distribution, the
two asymptotic branches of $P(t)$ have exponential relaxations,
whence one can extract two characteristic times: $\tau_-\sim57$ and
$\tau_+\sim 16$, associated to rare events.
\begin{figure}[h]
\centerline{  \resizebox{8cm}{!}{\includegraphics{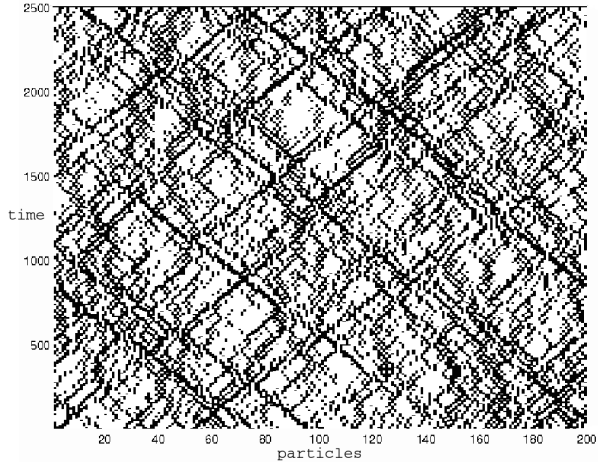}}}
\caption{Hypsometric plot of the energy density for the harmonic system: sites with an
  energy above (resp. below) the mean value are black (resp. white).}\label{hypso_V3}
\end{figure}
These times are the quantitative traduction of what the eyes
can observe in an hypsometric plot like that of fig \ref{hypso_V3}:
the sites experiencing long depletions of energy are related to rather
large white patches of fig. \ref{hypso_V3};  the quite large characteristic time
$\tau_-$  tells us that the occurence of large patches
of that kind are not so rare. On the contrary, a stagnant excess of
energy is much more unlikely due to the group velocity which
generically moves the wave crests; only for colliding wave crests can
occur a long stagnation, but this is rare. All these features
can be viewed as
emergent properties of the phonon dynamics with stochastic initial
conditions. The intrinsically rich nature of the persistence concept
probably prevents a theoretical calculation, despite the fact that the
system is integrable.

\subsection{Soft potential $V(x)=V_2(x)$}

\begin{figure}[h]
\centerline{  \resizebox{14cm}{!}{\includegraphics{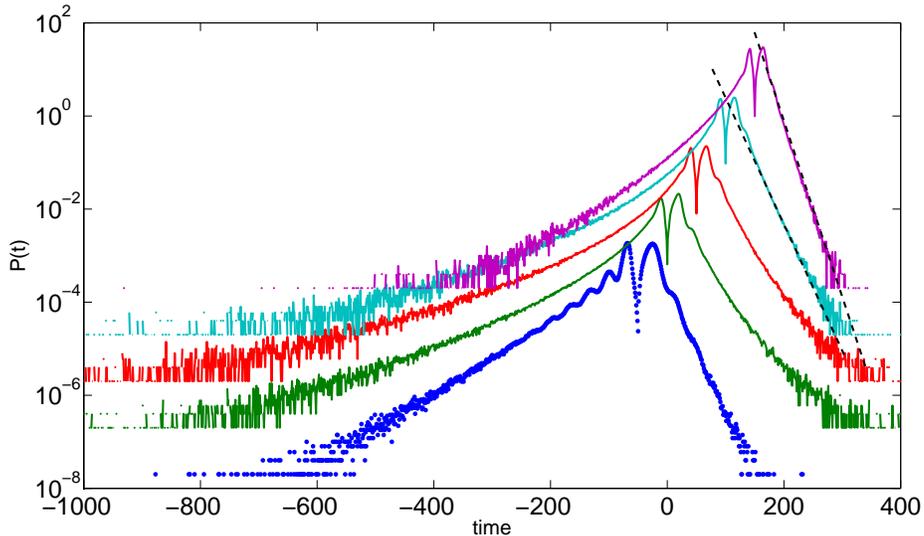}}}
\caption{Distribution $P(t)$ for the soft potential $V_2(x)$ at
  various temperatures. From bottom to top : harmonic potential
  (i.e. $V_2$ for $k_BT\rightarrow 0$),
  $k_BT=0.25$, $k_BT=0.5$, $k_BT=1.0$, $k_BT=2.0$. The curves were
  shifted in both directions for sake of clarity.}\label{TIPEE_V2}
\end{figure}

\begin{figure}[h]
\centerline{\resizebox{14cm}{!}{\includegraphics{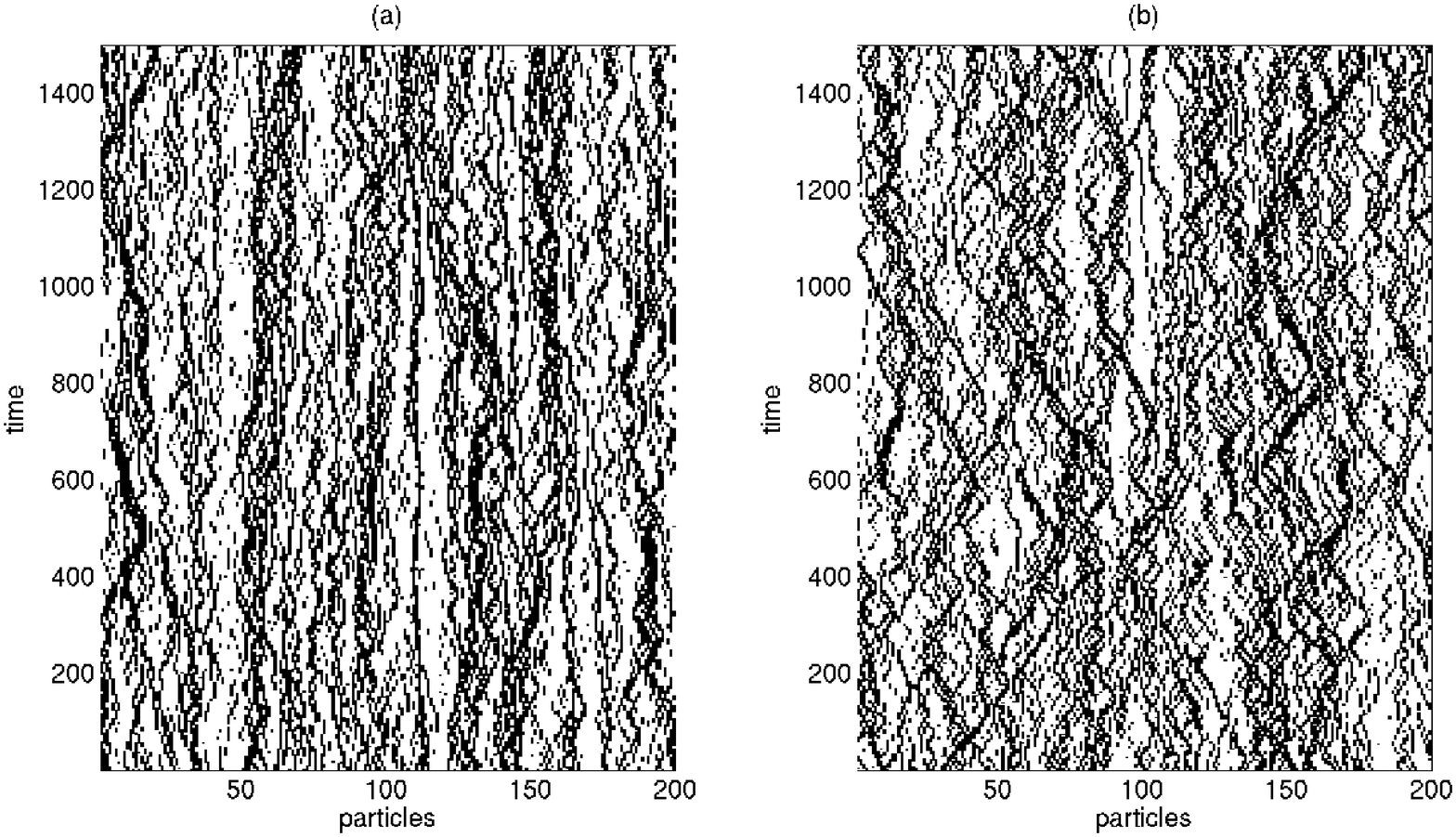}}}
\caption{Hypsometric plot of the energy density for the potential
  $V_2(x)$. Left: $k_BT=0.25$, right: $k_BT=2.0$. The temperature
  $k_BT=0.25$ corresponds to the minimum value of the diffusion coefficient.}\label{hypso_V2}
\end{figure}

For a soft potential like $V_2(x)$, moving breathers are
expected. They must lead to a TIPEE distribution not qualitatively
different from the preceding case. This is almost the case, as figure
\ref{TIPEE_V2}  shows. 
The distribution depend now on the
temperature, and the exponential tails are still
present except near the temperature $k_BT=0.25$. This temperature  corresponds to a minimum in the
diffusion coefficient, a minimum resulting from the simultaneous destruction of the
phonon structure by the nonlinearity and the excitations of rather
static breathers (see figure \ref{hypso_V2}, left panel). Actually, for the $V_2$ potential, the mobility of
breathers is an increasing function of their energy (or the
temperature). The static breathers perturbs the  picture of
mobile entities passing in a finite time through a particular site,
colliding into each other, etc\ldots, and this perturbation  leads to nonexponential
tails in the TIPEE distribution. 
These nonexponential tails anticipate
the behaviour of the potential $V_1(x)$, where static breathers are
the rule rather than the exception. Here, by increasing the
temperature, the thermal breathers become mobile (see
fig. \ref{hypso_V2} right panel) and the exponential
tails are restored in the distribution: this is indicated in figure
\ref{TIPEE_V2} by the dotted lines added to the curves for $k_BT=1$
and $k_BT=2$. On the contrary, the exponential behaviour is not
restored on the left wing of the distribution: the rare occurence of
lasting depletions is ``less rare'' than in the harmonic case,
probably due to the fact that the calm regions have frequencies
detuned with those of breathers, and thence can act as reflective
barriers for them, lengthening their lifetime. Such a
scenario can be clearly observed on the right panel of
fig. \ref{hypso_V2} (near particles numbered 100 and 180).

\subsection{Hard potential $V(x)=V_1(x)$}

\begin{figure}[h]
\centerline{\resizebox{14cm}{!}{\includegraphics{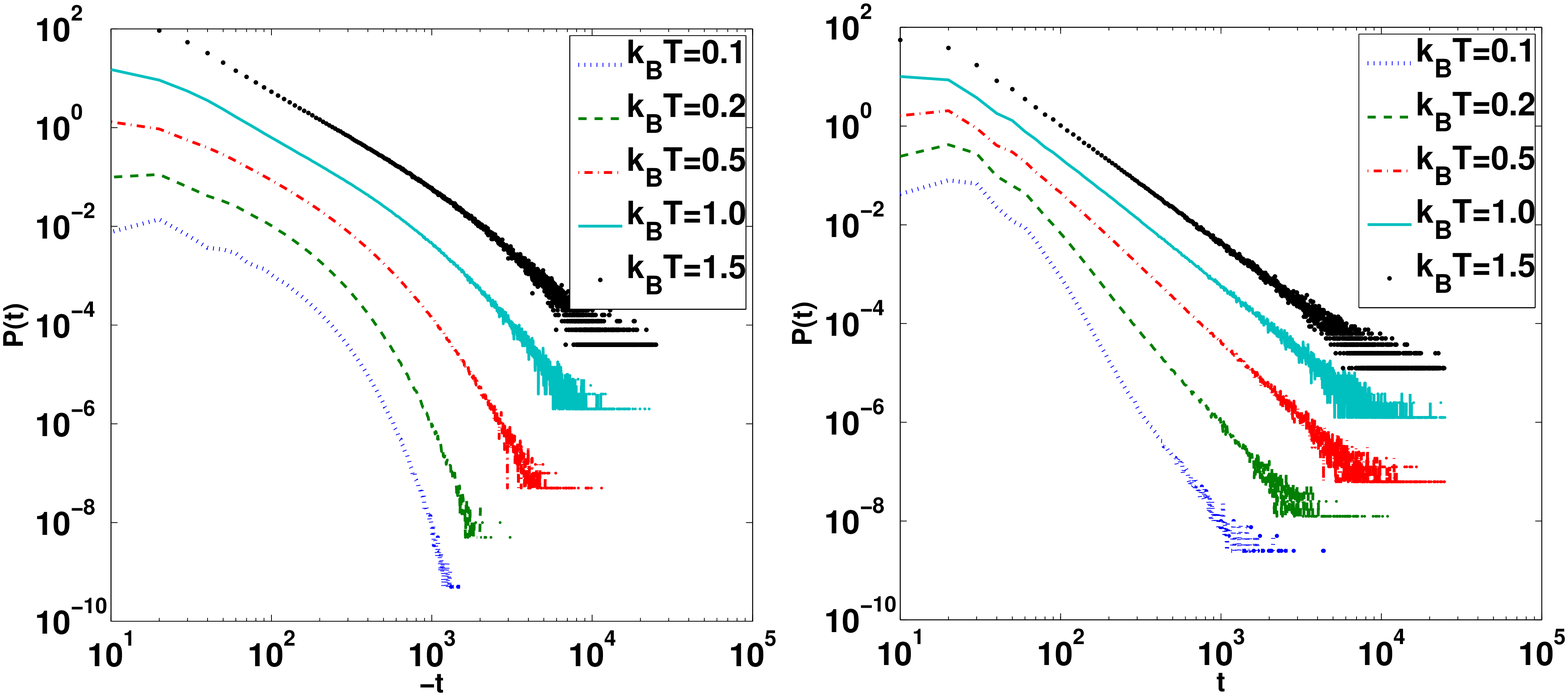}}}
\caption{Distribution $P(t)$ for the hard potential $V_1(x)=\demi x^2+\frac{1}{4}x^4$ at
  various temperatures. Left panel: negative times, right panel:
  positive times. The curves were
  shifted in vertical direction for sake of clarity.}\label{TIPEE_V1}
\end{figure}
For the hard potential $V_1(x)=\demi x^2+\frac{1}{4}x^4$, a remarkable
feature emerges for positive $t$ (see fig \ref{TIPEE_V1}, right
panel) : except for very low $T$, where the system resembles the
harmonic system, the asymptotic tail is clearly a powerlaw: no
characteristic time is associated to long lasting breathers, which
assume surprisingly this simple distribution law. Why this is so is
not easy to understand, as it is related to the local structure of the
dynamical phase space, but a phenomenological model can
be attempted. Let us assume that the energy $\eps$ of a given site obeys a simple
Fokker-Planck equation; owing to the equilibrium distribution
$P_{eq}(\eps)\propto \exp(-\be\eps)$ (we neglect the entropic
contribution of the phase degree of freedom), it reads
\begin{align}
  \pa_tp=\demi\frac{\pa}{\pa \eps}\left[B(\eps)[\be p+\pa_\eps p]\intvide\right]
\end{align}
where $\be=1/k_BT$ and $B(\eps)$ is an unknown diffusion
coefficient. This modelling is rather crude, as it disregards
completely the conserved character of the energy as well as the
extended nature of the system. The diffusion
coefficient $B(\eps)$ is a priori unknown, but we know that it is a
decreasing function of the energy. The simplest Ansatz we can make is
to choose $B(\eps)\propto\exp(-\nu\eps)$, where $\nu$ is a positive coefficient.
According to a procedure described in \cite{faragoeurophys}, one can
derive the asymptotic behaviour of the persistence distribution. In
that case, it yields an appealing result, namely a temperature
dependent powerlaw :
\begin{align}
  \mcal{P}_+(t)&\propto \frac{1}{t^{1+\be/\nu}}\\
\Longrightarrow P(t)&\propto \frac{1}{t^{3+\be/\nu}}\label{gollum}
\end{align}
Thus, the exponential Ansatz for $B(\eps)$ seems reasonable, but we
could also attempt a powerlaw Ansatz $B(\eps)\propto 1/\eps^\mu$. The
same demonstration from \cite{faragoeurophys} yields
\begin{align}
\mcal{P}_+(t)\propto t^{-\frac{\mu-1}{2\mu+2}}\times\exp[-(t/t_0)^{1/(1+\mu)}]
\end{align}
which is clearly not appropriate here.

\begin{figure}[h]
\centerline{\resizebox{7cm}{!}{\includegraphics{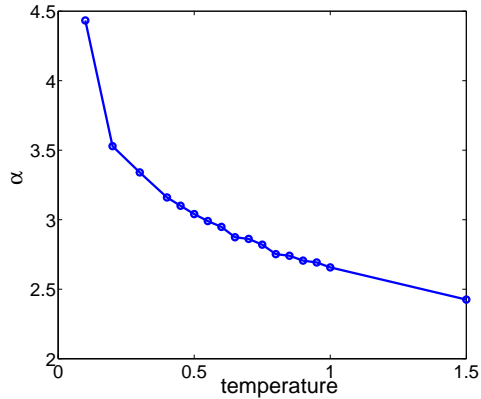}}}
\caption{Exponent $\alpha$ associated to $P(t)$ for the potential
  $V_1$. $\alpha$ is defined by $P(t)\propto 1/t^\al$ for large
  positive $t$}\label{fit_Pdet_exposant}
\end{figure}
So far so good, it seems from \myref{gollum}. Actually, the result is
not that good: to see that, we extract from the curves in figure
\ref{TIPEE_V1} (right panel) the exponents for each temperature. They
are plotted on figure \ref{fit_Pdet_exposant}, and it is clear that the exponent
is overestimated by this simple approach: the actual exponent is lower
than 3 for high $T$, which is not possible in the Fokker-Planck approach. The
positive point is that the temperature dependence is qualitatively
similar, i.e. the temperature lowers the exponent (the curve is however
not described by $C^t+C^{t'}/T$). One could argue the problem could be
solved by choosing a kernel $B$ decreasing much faster for high
$\eps$. But one can show that for $B(\eps)\propto \exp(-\nu
\eps^{1+\eta})$, one has asymptotically   $\mcal{P}_+(t)\propto 1/t$
irrespective of the values of $\nu$ and $\eta>0$.

A possible explanation for these  disagreements is to be
traced back to the fact that the Fokker-Planck approximation for the
energy diffusion is a markovian one, which is a rather poor
assumption for pinned thermal breathers in 1D : they are so to speak
squeezed in between two other breathers which act as reflecting barriers for
the mobile phonons and prevent a rapid decorrelation of the
noise. This too slow evolution of the thermal degrees of freedom slows
the degradation of the breathers and give them a longer
persistence. To conclude, the inability of that simple model to
quantitatively account for the phenomenology tells us that the giant
slowing down experienced by the system is not only due to an effective
diffusion coefficient in the energy space (the kernel $B(\eps)$) which
would decrease for increasing energy, but that the geometrical
constraints, squeezing in each breather between two other breathers
give a marked non markovian character to the dynamics and therefore plays a
prominent role to the overall slowing down. But, why the persistence
distribution of the breathers is powerlaw for model $V_1$ remains an
interesting unsolved question. 

\section{The power spectrum}

It is useful to have a look at the adimensionate power spectrum
$S(\om)$ defined by $\lan v_i^*(\om')v_i(\om)\ran=k_BT
S(\om)\de(\om-\om')$ where $v_i(\om)$ is the Fourier Transform of
$v_i(t)$. Roughly speaking, $S(\om)$ gives the frequency content of
the local thermal excitations, and in principle should bear a trace of
the thermal ``breathers''. 
\begin{figure}[h]
\centerline{\resizebox{9cm}{!}{\includegraphics{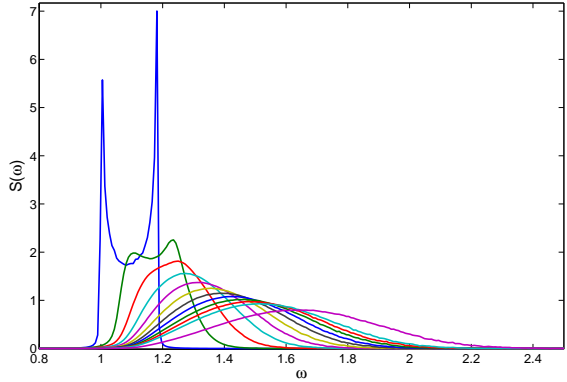}}}
\caption{Power spectrum $S(\om)$ for various temperatures : from left
  to right, harmonic system ($T\rightarrow 0$), $T=0.1(0.1)1.0$ and $T=1.5$.}\label{powsp:V1}
\end{figure}
The power spectrum for the potential $V_1$
is plotted in figure \ref{powsp:V1} for various temperatures (notice
that the curve for $T\sim 0$ corresponds to the harmonic potential $V_3$) . The
global shift of the spectra to the right for increasing temperature is
obviously due to the ``hard'' features of the potential: for a single
oscillator, the frequency increases with the particle energy. It is
remarkable that no blatant evidences of the breathers can be isolated,
like peaks growing outside the phonon band for instance, a fact
already noticed previously \cite{eleftheriouflachtsironis}. This is due to the fact that breathers of
various energies are stable without restriction above the upper bound
of the phonon band.

However, one expects that for temperatures high enough, at least the
right asymptotic of the spectra (the part rightwards the maximum)
accounts for breathers. As the distribution of breathers, or more
precisely, the distribution of their persistence, is powerlaw, and retains this shape for various temperatures, one expects
these right wings to be more or less universal.
\begin{figure}[h]
\centerline{\resizebox{7cm}{!}{\includegraphics{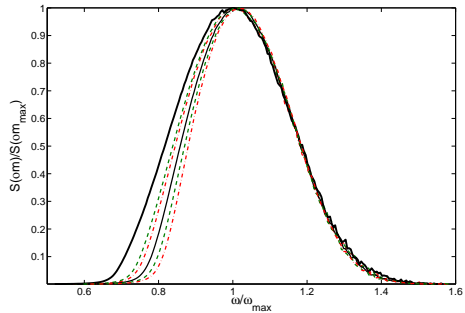}}}
\caption{Rescaled power spectra $S(\om)/S(\om_{max})$ as a function of
  $\om/\om_{max}$ for various temperatures : from left
  to right, $T=1.5,1.0,0.9,0.8,0.7,0.6$}\label{powsp_ren:V1}
\end{figure}
 Figure
\ref{powsp_ren:V1} represents various power spectra from $k_BT=0.6$ to
$k_BT=1.5$ where the axis were stretched so that the maxima of the
curves coincide at $(1,1)$  (we allowed also an extra small horizontal translation
of at most $5.10^{-2}$ to allow for the numerical uncertainty of the position
of the maximum). The figure shows clearly that the right
side of the curve can be described by the same master curve, provided
the temperature is not too low (for $k_BT\leq0.5$ we observed a
noticeable departure from the collapsed right branch).  On the
contrary, the left branches do not collapse on a single curve but
stretch to the left for increasing temperatures. This is related to
the fact that the ratio $\Delta \om/\om_{max}$, where $\Delta \om$ is
the frequency width of $S(\om)$ and $\om_{max}$ its maximum, increases
with the temperature.

\medskip

\begin{figure}[h]
\centerline{\resizebox{14cm}{!}{\includegraphics{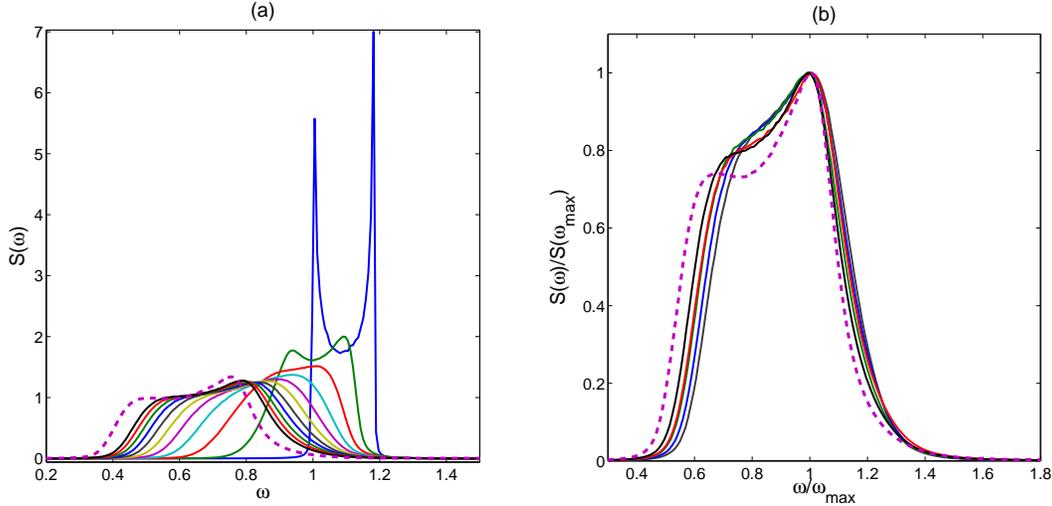}}}
\caption{Natural (a) and rescaled (b) power spectra for the potential
  $V_2$ and for various temperatures. For (a) : from right
  to left, $k_BT=0(0.1)1.0,1.5$. For (b) $k_BT=0.6(0.1)1.0,1.5$. For
  sake of clarity, the curve corresponding to $k_BT=1.5$ has been
  dashed in both plots.}\label{powsp:V2}
\end{figure}

It is not uninteresting to have a look at the power spectra associated
with the soft potential $V_2$ (see fig \ref{powsp:V2}).
As expected, the power spectra shift toward low frequencies when
temperature is increased. Moreover, as already noticed with the
persistence distribution, high temperatures restore an effective
phonon behaviour, which is embodied here by the bimodal structure of
the curve, shaded for intermediate $k_BT$ but restored for higher
$k_BT$. Finally, the same rescaling as in figure \ref{powsp_ren:V1}
(in fig. \ref{powsp:V2} (b)) does not highlight any collapse as for
$V_1$: this confirms indirectly the intimate relation between the
collapse of fig. \ref{powsp_ren:V1} with the thermal static breathers.

\section{Anomalous or normal diffusion ?}

\begin{figure}[h]
  \centerline{\resizebox{7cm}{!}{\includegraphics{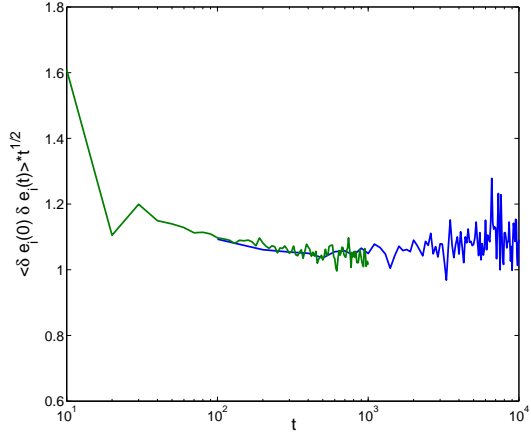}}}
\caption{For the potential $V_2$, (energy density autocorrelation
  function)x(square root of time) as a function of time. The two
  curves probe the same system at different time ranges ($k_BT=1$).}\label{diff_logcosh}
\end{figure}

\begin{figure}[h]
  \centerline{\resizebox{7cm}{!}{\includegraphics{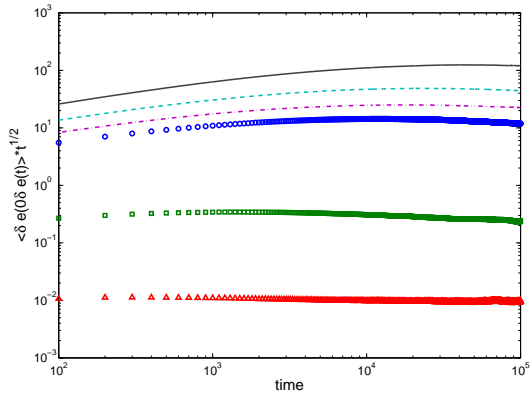}}}
\caption{For the potential $V_1$, (energy density autocorrelation
  function)x(square root of time) as a function of time. The different
  curves correspond to temperatures $k_BT$=0.1 (triangles), 0.3
  (square), 1.0 (circles), 1.2 (dash-dot), 1.5 (dashed), 2.0 (solid).
  }\label{diff_x4}
\end{figure}

A natural question raises when considering the system with the
potential $V_1$. The nucleation of pinned breathers which besides
reflects efficiently the phonons (by the frequency detuning),
affects deeply the diffusion properties of the system; does it induce
a \textit{qualitative} change toward an anomalous subdiffusive
behaviour ? The numerical results as those of \cite{tsironisaubry}
suggest that it is indeed the case, but their numerical experiments
are to some extent ``extremely'' out of equilibrium near the
boundaries, leading to a enormous detuning; as far as I know, the
normality of the energy diffusion has not been tested at equilibrium.
For a system having a normal energy diffusion, the autocorrelation of
the energy density must behave $\propto 1/\sqrt{t}$ at large
times. This is for instance the case for the system with the potential
$V_2$, as shown in figure \ref{diff_logcosh}; in that figure,
the autocorrelation of energy $\lan \de e_i(0)\de e_i(t)\ran$ is multiplied by $\sqrt{t}$, in order to
get a plateau for a normal diffusive behaviour. In figure
\ref{diff_logcosh}, this plateau is unmistakable, despite
unavoidable fluctuations due to a lack of statistical accuracy. Now
let us consider the system with the hard potential $V_1$. The figure
\ref{diff_x4} is the same as fig. \ref{diff_logcosh}, except that the
potential is $V_1$ (and the plot is log-log).
Clearly, for low temperatures, the system displays a normal diffusive
behaviour. For higher temperatures, the slowing down induced by the
thermal breathers makes the curves  increase higher the higher
$k_BT$, but for each curve computed, this increase ceases and the curves
have a maximum. It would be difficult numerically to probe these
curves for longer times, but we think that the figure \ref{diff_x4}
shows convincingly that the ultimate regime of relaxation for the
system is a \textit{normal} one: obviously, the curves cannot decrease
to zero: it would lead to a superdiffusive behaviour which is
physically absurd; thus, it is most probable that these curves reach
eventually a plateau regime, but it is worth stressing that this
ultimate regime is extremely remote for high temperatures: for
$k_BT=1.5$ or $2$, at $t=10^5$, the system has not reached its
asymptotic regime. Thus, in that kind of system, the pinned thermal
breathers do not modify the very nature of the energy diffusion, but
slow down enormously the diffusion: as a result, the signature of
thermal breathers one can get not only in the very small values of the
diffusion coefficients, but also in the very long transients
observable in the correlation functions of the energy density. But
this study indicates that it is not correct to see a system with
breathers as a vitreous system, as at equilibrium there is no
indication that anomalous diffusion takes place, not to mention
ergodicity breaking.

\section{Breather lifetime}

As we have already stressed,  ``breathers'' as numerable entities do
not exist for generic states of the system. However, for the system
with the hard potential $V_1$, the hypsometric plots
\cite{tsironisaubry} show clearly that
for not too low $k_BT$, the eye can separate without much ambiguity
different excitations, almost static, which it is tantalising to term
``breathers''. Thus, in that case, a phenomenological approach based
on an effective population of breathers is not meaningless, and it is
interesting to confront that approach to the results we have got for
the persistence. In the following we consider only the model $V_1$.

We assume that the energy into the system is dispatched amongst $N_b$
breathers. The distribution of the free energies is taken $p(E)\propto
\exp(-\be E)$, what is slightly uncorrect, as the entropic degrees of
freedom (associated to the phase coordinate of the breathers) are
neglected; but for sake of simplicity, we retain this exponential law.
In principle, the number $N_b$ can be related to the number $N$ of particles via the
formula
$N_b\lan E\ran=N\lan e_i\ran$, but for $k\rightarrow 0$, one has
$N_b\approx N$. 

These $N_b$ breathers are essentially static excitations with long
lifetimes. We can assume that the right tail of the distribution
$P(t)$ is entirely constructed by these breathers. As the lifetime of
a breather is obviously an increasing function of its energy, we can
write $P(t)dt\propto p(E)dE$. Thus, for high energies,
$dt/t^\al\propto \exp(-\be E)dE$, whence one deduces the lifetime of a
breather as a function of its energy :
\begin{align}
  t(E)\propto \exp\left(\frac{\be E}{\al-1}\right)
\end{align}
These lifetimes are thus exponentially related to the energies, what
is reminiscent of an Arrhenius law. It is probably quite bold or even
incorrect to surmise that the decay of a breather is an activated
process, as it is clear that no Peierls-Nabarro barrier does exist for
breathers; anyway, a barrier can be sometimes entirely due to
entropic effects, namely, the shift of a breather from one site to an
adjacent one (what is, owing to the definition of the persistence,
equivalent to the ``death'' of a breather at his starting site) could
occur without energy changes but through a succession of different
states so tailored that it makes an effective entropic barrier. Thus,
if we follow this line, the effective depth of the local well where the
system oscillates when a breather of energy $E$ exists would be given
by $E/(\al-1)$ : the depth is proportionnal to the energy of the
excitation but lower than it; it is temperature dependent, and tends
to increase with increasing $T$ ($E$ being fixed). 

To proceed further, it would be necessary to study carefully the
multidimensional phase space of these systems (and the corresponding
dynamics), to see if the theory of dynamical systems can give more
consistency and support to these phenomenological remarks.

\section{Conclusion}

In this paper, we have studied different Klein-Gordon chains at
thermal equilibrium. These systems possess breather solutions of the
Hamiltonian dynamics, which can be considered as periodic excitations
at zero temperature. At $k_BT\neq 0$, the concept of breather is void
in principle, but transient local maxima of energies which behave like
individual items can be observed in these systems near the
anticontinuum limit, which are termed thermal breathers by
extension. We have shown that the concept of persistence is well
adapted to study these objects when they are firmly pinned as in most
of Klein-Gordon chains with hard nonlinear potentials. Interestingly
enough, the persistence distributions are asymptotically powerlaw,
with an exponent decreasing with the temperature. This remarkable
feature has been shown to be intimately related to both the geometrical structure of the
phase space and the non markovian character of the local
dynamics. But the very explanation of this ``multifractal''
distribution in terms of dynamical systems remains an open issue.
Moreover, we have also shown that this particularly simple distribution of
breathers of high energies has a translation in the power spectra :
using axis rescaling to superimpose the maxima, it turned out that the
high frequency tails of the distributions coincide exactly, witnessing
that a certain universality ---some would term it perhaps
``self-organized criticality''--- is at work in the dynamics of breathers,
provided the temperature is high enough to have destroyed completely
the phonon structure.

This work raise some other issues: for discrete nonlinear systems in
2d or 3d, what tell the persistence or TIPEE distributions ? Clearly,
the topology is much less a constraint, and the slowing down of the
energy diffusion is less pronounced; but it would be interesting to
test if the thermal breathers affect sensibly the persistence
distributions, yielding for instance nonexponential asymptotic
tails. Another issue is related to systems like model $V_2$: clearly
for systems where the thermal
``breathers'' are mobile, following the persistence of a given site is
not the ideal way of having signatures of breathers. The natural idea
would be to define an extended persistence concept, allowing to follow
a positive fluctuation of energy from one site to an adjacent
one. This idea is appealling, but hides a certain number of technical
difficulties related to the fact that instead of following a site, one
follows a positive fluctuation: one implicitely tries to extract a
breather distribution with all the ambiguities associated to it.

\section{Appendix: Relation between TIPEE and persistence}

Let us consider a large time interval $[0,T]$, when a stochastic
process $Y(t)=X(t)-\lan X\ran$ perform a number of sign flips. The
time interval divides in $N\gg 1$ time intervals
$\tau_1,\ldots,\tau_N$ where $Y$ keeps a constant sign, and $\tau_i$
is taken negative if $Y(t)<0$ on that time interval. The repartition
function of the $\tau_i$ is nothing but the TIPEE (Time Intervals of
Persistent Energy Excess) $P(\tau)$. We have in particular 
\begin{align}
  \int_{-\infty}^\infty d\tau P(\tau)=1
\end{align}
The persistance above zero for $Y$ is defined as the probability
$\mcal{P}_+(\tau)$ that ``$Y(t)$ stays above zero for a time longer than
$\tau$, knowing it started from a positive value''. The probability
that it started from a positive value is just $\left(\int_0^\infty dt\
tP(t)\right)/\left(\int_{-\infty}^\infty dt\ tP(t)\right)$. The
probability that $Y(t)$ stays above zero for a time $>\tau$ and it
started from a positive value, is associated to the chance we have to
pick up at random a time in $[0,T]$, such that we are in a positive
interval of length greater than $\tau$, and far enough from the end
such that the condition is fulfilled. This probability is thus given
by
\begin{align}
  \int_\tau^\infty
  d\tau'\left(\frac{P(\tau')\tau'}{\int_{-\infty}^\infty d\tau''|\tau''|P(\tau'')}\right)\times\frac{\tau'-\tau}{\tau'}
\end{align}
whence we get
\begin{align}
  \mcal{P}_+(\tau)=\int_\tau^\infty d\tau'\
  P(\tau')(\tau'-\tau)\left/\int_0^\infty d\tau' P(\tau')\tau'\right.
\end{align}
We verify that $\mcal{P}_+(0)=1$. The corresponding formula for
$\mcal{P}_-(\tau)$ is readily obtained by changing $\tau$ by $-\tau$
in $P$.


\begin{thebibliography}{99}
  \bibitem{flachwillis}
S. Flach and C.R. Willis, \textit{Physics Reports},{\bf 295}, 181-264
(1998).
\bibitem{aubry60} S. Aubry, \textit{Physica D},{\bf 216},1-30 (2006).
\bibitem{mackayaubry}R.S. MacKay and S. Aubry, \textit{Nonlinearity},{\bf 7},1623-1643 (1994).
\bibitem{tsironisaubry} G.P. Tsironis and S. Aubry,
  \textit{Phys. Rev. Lett.},{\bf  77} 5225-5228 (1996).
\bibitem{lindenberg} R.Reigada, A.H.Romero, A.Sarmiento, K.Lindenberg,
  \textit{J.Chem.Phys.},{\bf 111}, 1373-1384 (1999).
\bibitem{eleftheriou} M. Eleftheriou and S.Flach, \textit{Physica D},
  {\bf 202}, 142-154 (2005).
\bibitem{ivanchenko} M.V.Ivanchenko, O.I.Kanakov, V.D.Shalfeev,
  S.Flach, \textit{Physica D}, {\bf 198}, 120-135 (2004).
\bibitem{eleftheriouflachtsironis}
M. Eleftheriou, S. Flach and G. P. Tsironis, \textit{Physica D}, {\bf
  186},20-26 (2003).
\bibitem{majumdar}S.Majumdar, \textit{Curr. Sci.}, {\bf 77} 370
  (1999).cond-mat/9907407
\bibitem{faragoeurophys} J.Farago, \textit{Europhys.Lett.},{\bf 52},
  379 (2000).
 \end{thebibliography}
\end{document}